\documentstyle[12pt,prl,aps,preprint]{revtex}
\begin{document}
\renewcommand{\baselinestretch}{0.3}
\draft
\title{A simple proof of the unconditional security of\\
 quantum key
distribution}
\author{Hoi-Kwong Lo}
\address{Hewlett-Packard Labs,
Filton Road, Stoke Gifford, Bristol, UK, BS34 8QZ.\\
 E-mail: hkl@hplb.hpl.hp.com}

\date{\today}
\maketitle
\begin{abstract}
Quantum key distribution is the most well-known application of quantum
cryptography. Previous proposed proofs of security of quantum
key distribution contain various technical subtleties. Here, a
conceptually simpler
proof of security of quantum key distribution is presented.
The new insight is the invariance of
the error rate of a teleportation
channel:
We show that the error rate of a teleportation channel is independent of
the signals being transmitted. This is because
the non-trivial error patterns are permuted under teleportation.
This new insight is combined with the recently proposed quantum to
classical reduction theorem.
Our result shows that assuming that Alice and Bob have fault-tolerant quantum
computers, quantum key distribution can be made unconditionally
secure over arbitrarily long distances even against the most general type of
eavesdropping attacks and in the presence of all types of
noises.

\end{abstract}
\narrowtext
\tighten
\section{Introduction}
\label{Intro}
Perfectly secure communication between two users can be achieved
if they share beforehand a common random string of
numbers (a key).
A big problem in conventional cryptography is
the key distribution problem: In classical physics, there is nothing to
prevent an eavesdropper from monitoring the key distribution
channel passively, without being caught by the legitimate users.
Quantum key distribution (QKD) \cite{BB84,Ekert}
has been proposed as a new solution to the
key distribution problem.
In quantum mechanics, there is a well-known
``quantum no-cloning theorem'' which states that it is
impossible for anyone (including an eavesdropper) to make a perfect
copy of an unknown quantum state \cite{Dieks,WZ}.
Therefore, it is generally thought
that eavesdropping on a quantum channel will almost surely produce
detectable disturbances.

\subsection{Prior work on security of QKD}
``The most important
question in quantum cryptography is to determine how
secure it really is.'' (p. 16 of \cite{BC})
Indeed, there have been many investigations on the issue of
security of QKD. Most analyses have dealt with restricted
classes of attacks such as single-particle
eavesdropping strategies (For a review, see, for example, \cite{Lobook}.),
and also the so-called collective attacks \cite{scollect1,scollect2},
where Eve brings each signal particle into interaction with
a separate probe, and after hearing the authenticated public
discussion between Alice and Bob, measures all the probes together.
More recently, the most general type of attacks have been considered.
There have been a number of
proposed proofs of the {\it unconditional} security of
QKD \cite{sdeutsch,LCqkd,smayers1,smayers2,smayers3}
based on the laws of quantum mechanics.
Note that one should also consider problems of imperfect
sources, imperfect measuring devices and noisy channels
employed by Alice and Bob.

\subsubsection{Why is a proof of security of QKD so difficult?}
\label{why}
There are many types of eavesdropping strategies.
One could imagine that Eve has a quantum computer.
In the most general eavesdropping strategy, Eve regards the
whole sequence of quantum signals as a single entity.
She couples this entity with her probe and then
evolves the combined system using a unitary transformation
of her choice. Finally, she
sends a subsystem to the user(s) and keeps the rest
for eavesdropping purposes. Notice that
Eve can choose any unitary transformation she likes and
yet a secure QKD scheme must defeat all of them.
Two major difficulties are expected in a proof of security
of QKD. First, Eve tries to evade detection by attributing
noises caused by her eavesdropping attack to normal
transmission noise. Second, owing to the subtle quantum
correlations between Eve and the users, a na\"{\i}ve application
of classical arguments may be fallacious. Indeed, there
is a well-known paradox---Einstein-Podolsky-Rosen paradox
\cite{EPR}---which
illustrates clearly the general failing of
na\"{\i}ve classical arguments in quantum mechanics.

\subsubsection{Two alternative approaches to proving security}
\label{ss:two}
Roughly speaking, there are two main alternative approaches to
proving the unconditional security of QKD.
The first approach deals with the most well-known
QKD scheme BB84 proposed by Bennett and Brassard \cite{BB84}.
The advantage of this approach is that it does not
require the employment of quantum computers by Alice and Bob.
However, all versions of current proposed proofs of unconditional security
based on this
approach require the assumption of a perfect photon source
\cite{smayers3}. (Earlier versions of \cite{smayers3}
have appeared as \cite{smayers1,smayers2} but
they are less definite.)
Given that a perfect photon source is beyond current
technology, proofs based on the first approach (just like those
based on the second approach) cannot be directly applied to
real-life experiments. See also \cite{Mayao}.
The second approach deals with QKD schemes that
employ the subtle quantum mechanical correlations---known
as ``entanglement''--which have no classical analog.
This approach was first suggested in \cite{sdeutsch},
which, however, assumes perfect quantum
devices. A more recent paper \cite{LCqkd}
addresses this issue of imperfect devices using the idea of
fault-tolerant quantum computation and quantum repeaters
(i.e., relay stations) \cite{Dur}. It also derives a
rigorous bound on Eve's information
under the assumption of reliable local quantum computations.
Note that the second approach requires Alice and Bob to possess quantum
computers, which are well beyond current
technology. However, the second approach,
as rigorously developed in \cite{LCqkd}, has a number of advantages.
First, it extends the range of secure QKD to arbitrarily long
distances even with insecure ``quantum repeaters''
(i.e., relay stations).
In contrast, such an extension with the first approach will
require perfectly secure quantum repeaters.
Second, when implemented over a noisy channel, QKD schemes
based on the second approach tend to tolerate a larger
error rate. Third, a proof of security and the tradeoff between
noise and key rate are much easier to work out in the second
approach. Fourth, the second approach is conceptually simpler.
Finally, some of the techniques developed in the second
approach have widespread applications. Indeed, it is plausible
that some of those techniques, when properly generalized, can be
applied to the first approach.

\subsection{Significance of Our results}
It has to be said that all previously proposed proofs of security of
QKD involve various technical subtleties.
Here we present a simple proof of
the unconditional security of QKD.
The proof, based on the
second approach, not only enjoys all the fundamental advantages mentioned
above of the recently
proposed proof\cite{LCqkd},
but also is conceptually simpler.

Besides, our proof gives us an extremely interesting new
insight on the well-known ``teleportation'' channel \cite{tele}: With a
classical random sampling method,
one can assign a set of {\it classical} probabilities to the various
error pattern of a {\it quantum} teleportation channel. Besides, the error
rate (the probability of having a non-trivial error pattern)
for each signal is independent of the identity of the signal being
transmitted. This is highly non-trivial because,
as noted in subsubsection~\ref{why}, the well-known
Einstein-Podolsky-Rosen paradox demonstrates that applications of
classical arguments to a quantum problem often lead to fallacies \cite{EPR}.

Another potential advantage of our proof is that, with
imperfect local quantum computations, it is
probable that a longer key can be generated by the
current scheme with the same quantum channel.

\section{Security requirement and ideas towards a proof}
{\bf Definition}: A QKD scheme is said to be unconditionally secure if,
for any security parameters $k, l > 0$ chosen by Alice and Bob,
they can follow the protocol and
construct a verification test such that, for
{\it any} eavesdropping attack by Eve that will pass the test
with a non-negligible amount of probability,
i.e., more than $e^{-k}$, the two following conditions are
satisfied:
(i) Eve's mutual information with the final key is always
negligible, i.e., less than $e^{-l}$ and (ii) the final key is, indeed,
essentially random.

{\it Remark}: The security parameters $k$ and $l$ depend on
how hard Alice and Bob are willing to work towards
perfect security (e.g., the
size of the messages exchanged between Alice and
Bob and the number of rounds of
authentication between them) and are,
at least in principle, computable from a protocol.

\subsection{A simple idea, its problems and our solution}
Consider the following simple idea of proof of security of QKD.
Alice prepares $r$ quantum signals and encodes their state
into a quantum error correcting code (QECC) (see, for example,
\cite{BDSW}) of length $n$ which
corrects say $t$ errors.
In addition, she also prepares $m$ other quantum signals
which will be used as test signals.
She then {\it randomly} permutes the $N=n+m$ signals
and sends them to Bob via a noisy channel
controlled by an eavesdropper.
Bob publicly announces that he has received all the $N$ signals
from Alice. Upon Bob's confirmation of the receipt,
Alice publicly announces the location of the
$m$ test signals and their specific state.
Now, Bob measures the $m$ test signals and computes their
error rate, $e_1$. Using the error rate $e_1$, Alice and Bob
apply classical random sampling theory in statistics
to establish confidence levels
for the error rate of the $n$ remaining (i.e., untested)
signals and, hence, produce a probabilistic
bound on the amount of eavesdropper's information on the
encoded $r$ quantum signals. [The point is that,
unless there are more than $t$ errors in the QECC, Eve
knows absolutely nothing about the encoded state.]
If Alice and Bob are satisfied with the
degree of security, they measure the $r$ quantum
signals to generate an $r$-bit key.

This raw idea looks simple, but it is essentially classical.
It will work if the following three requirements are satisfied.
(1) Each error pattern
can be assigned with a classical probability;
(2) Error rate of the signals are independent of
the actual signals being transmitted
(i.e., Eve cannot somehow change a non-trivial error
operator to a trivial one depending on which signals are transmitted);
(3) The quantum error correction and key generation can be
done fault-tolerantly.

Since applications of
classical arguments could be fallacious,
it would be na\"{\i}ve to assign a probability
distribution to the set of error patterns without a rigorous
mathematical justification.
In fact, rather disappointingly,
we are unable to establish requirements (1) and (2) for the most
general quantum channel.

Nonetheless, we manage to
complete our proof of security of QKD by the following line of arguments.
We notice that requirement (1) has already been established
in \cite{LCqkd} for the special
case of the transmission of some standard states
(halves of so-called EPR pairs).
Moreover, it is well-known in quantum information theory
that the transmission of any general quantum state can be
reduced to that of the standard state and classical
communication via a process called {\it teleportation} \cite{tele},
(which will be discussed in subsection~\ref{ss:tele}).

Our line of attack is, thus, to establish requirements
(1) and (2) for the special case of  a teleportation channel
only.
In other words, we show that, by using teleportation to
transmit quantum states through a noisy quantum channel (which
may be controlled by an eavesdropper), the error rate
[i.e., the probability of having a non-trivial
error operator (or Pauli matrix) acting on the transmitted signal,
as can be estimated by a classical random sampling
procedure] is independent
of the quantum state being transmitted. This invariance result
ensures that, for a quantum teleportation channel, even
an ingenious eavesdropper
cannot change its underlying error rate and make it dependent
on the identity of
the quantum signals being transmitted.
This new insight of ours---the ``invariance of the
error rate of a quantum teleportation
channel''---will be stated as Proposition~5 and discussed
in subsequent
sections.

\subsection{Einstein-Podolsky-Rosen pairs}

Readers who are unfamiliar with quantum
information should
refer to appendix~\ref{physics}.
One can measure a quantum bit (or qubit) along any direction
and each measurement can give two possible outcomes.
An Einstein-Podolsky-Rosen pair of
qubits has the following interesting property. If two members of
an EPR pair are measured along {\it any} common axis, each member will give
a random outcome, and yet, the outcomes of the two members will
always be anti-parallel. This is so even when the two members
are distantly separated. Such an action at a distance is at the core
of the Einstein-Podolsky-Rosen paradox and it defies any simple
classical explanation.

Now, if two persons, Alice and Bob, share $R$ EPR
pairs, they can generate a common random string of number (an $R$-bit key)
by measuring each member along some common axis.
The laws of quantum mechanics guarantees that, provided that
the $R$ pairs are of almost perfect fidelity, the key
generated will be almost perfectly random and that
Eve will have a negligible amount of information on its value.
In fact, we have

{\bf Lemma~1}: (Note 28 of \cite{LCqkd})
If Alice and Bob share $R$ EPR pairs of {\it fidelity} at
least $ 1 - 2^{-k}$, for a sufficiently large $k$, and
they generate an $R$-bit key by measuring these pairs along
any common axis, then Eve's mutual information on the final key will
be bounded by $2^{-c} + 2^{O(-2k)}$ where $c= k - log_2
\left[ 2R + k + ( 1 / \log_e 2 ) \right]$.

{\bf Proof}: In supplementary material of \cite{LCqkd}.

So, the Holy Grail of the second approach to
secure QKD is to construct
a scheme for distributing $R$ almost perfect EPR pairs even in the
presence of noises and Eve.

\section{Quantum to Classical Reduction Theorem}
\subsection{Theory}
A proof of security of QKD can be simplified greatly if one
can apply well-known powerful techniques in classical
probability theory and statistical theory to the problem.
However, as noted in subsection \ref{why},
applications of classical arguments to a quantum problem
often lead to fallacies.
A key ingredient of our current proof is, therefore, a
quantum to classical reduction theorem
proven in \cite{LCqkd}, which
justifies the usage of classical arguments.

Let us recapituate this quantum to classical reduction theorem
from the viewpoint of ``commuting observables'':
Conceptually, classical arguments work because all the
observables $ O_i$'s under consideration are diagonal with
respect to a {\it single} basis, which we shall call $\cal B$.
More concretely, let $M$ be the observable that represents
the complete von Neumann measurement along the basis $\cal B$.
Since $O_i$'s and $M$ are all diagonal with respect to the
basis $\cal B$, they clearly commute with one another.
Therefore, the measurement $M$ along basis $\cal B$ will
in no way change the outcome of subsequent measurements $O_i$'s.
Without loss of generality, we can imagine that such a measurement
$M$ is always performed before the measurement of subsequent $O_i$'s.
Consequently, the initial state is always a classical mixture of
eigenstates of $M$ and, hence, classical arguments carry
over directly to a quantum problem. In this sense,
the quantum problem has a classical interpretation.\footnote{This
quantum to classical reduction theorem is rather subtle. First,
the observables $O_i$'s under consideration
are coarse-grained observables (i.e. observables with
degenerate eigenvalues), rather than fine-grained ones (i.e.
observables with non-degenerate eigenvalues). It is {\it a priori}
surprising that coarse-graining as a mathematical technique
will give a classical interpretation to a quantum problem.
Second, the
eigenstates of $M$ employed in \cite{LCqkd} are, in fact,
the so-called Bell states (see
subsection~\ref{application} and appendix~\ref{a:Bell}),
which exhibit non-local
quantum mechanical correlations. It is {\it a prior}
surprising that such a non-local (or quantum mechanical)
Bell basis can have a classical interpretation.}
Mathematically, this quantum to classical reduction theorem
can be stated as the following theorem.

{\bf Theorem~2}: \cite{LCqkd} Consider a mixed quantum
state described by $\rho$ and a set of one-dimensional
non-commuting projection operators $Q_j$ on it. Suppose
there exists a complete set of coarse-grained
observables $O_i$ of $Q_j$ such that all the $O_i$'s commute
with one another. [Here, by coarse-graining,
one means that each $O_i$ can be written as a sum of
a set of {\it orthogonal} projectors $Q_j$ and
by completeness, one means that
$\sum_i O_i = I$.] Let us consider a complete
von Neumann measurement $M$ which commutes with all $O_i$.
[Because of the commutativity of $O_i$'s, such $M$ must exist.]
Let $| v_k \rangle$ be the basis vectors of $M$.
Then, Theorem~2 says that, for all $i$, we have
\begin{equation}
{\rm Tr} \left( O_i \rho \right) 
={\rm Tr} \left( O_i \sum_k | v_k \rangle \langle v_k|  \rho
| v_k \rangle \langle v_k| \right).
\label{e:theorem2}
\end{equation}

{\it Remark}: Physically, Theorem~2 says that the probability of
all the coarse-grained outcome $O_i$'s are unchanged by a prior
complete von Neumann measurement $M$.
The full power of Theorem~2,
will be demonstrated in Propostion~3.

{\bf Proof}: Sketch. By construction, for each $O_i$
there exist a coefficient $\lambda_i$ and a set $K_i$ such that
$O_i = \lambda_i \sum_{l \in K_i} | v_l \rangle \langle v_l|$.
From the definition of ${\rm Tr} A $ as $\sum_m
\langle v_m | A| v_m \rangle$, it is now a simple exercise to establish
Eq.~(\ref{e:theorem2}).

\subsection{Application to random sampling}
\label{application}
Consider the following example
(example (i) on p. 2054 of \cite{LCqkd}).
Suppose two distant observers, Alice and Bob, share a large number,
say $N$, pairs
of qubits, which may be prepared by Eve.
Those pairs may, thus, be entangled with one another in
an arbitrary manner and also with the external
universe, for example, an ancilla prepared by Eve.
How can Alice and Bob estimate the number of
singlets in those $N$ pairs? (By the number of singlets, here we
mean the expected number of ``yes'' answers if a singlet-or-not
measurement were made on each pair individually.)

The solution is the following random sampling procedure and proposition.

{\it Procedure}: Suppose Alice and Bob randomly pick $m$ of the $N$ pairs
and, for each pair, choose randomly one of the three
($x$, $y$ and $z$) axes and measure the two members
along it. They publicly announce their outcomes.
Let $k$ be the number of anti-parallel outcomes
obtained in this random sampling procedure.

{\bf Proposition~3}: (in Section VI of supplementary material of
\cite{LCqkd}) The fraction of singlets, $f_s$, in the $N$ pairs
can be estimated as $(3k -m)/2m$. Furthermore, confidence
levels can be deduced from classical statistical theory for
a finite population (of $N$ objects).

{\bf Proof}: A direct application of Theorem~2.
Let us order the $N$ pairs.
Consider, for the $i$-th pair, the projection operations
$P^i_{\parallel,a}$ and $P^i_{{\rm anti}-\parallel, a}$ for the
two coarse-grained outcomes (parallel and anti-parallel) of the measurements
on the two members of the pair
along the $a$ axis where $a = x, y$ or $z$.
A simple but rather important
observation is the
following: each of these projection operators can be
mathematically
re-written as linear combination of projection operators along
a single basis, namely Bell basis. (See appendix~\ref{a:Bell} for
details.) A basis for $N$ ordered pairs of qubits (what
we shall call $N$-bell basis)
consists of products of Bell basis vectors,
each of which is described by a
$2N$-bit string.
Now, let us consider the operator $M_B$ that represents
the action of a complete von Neumann measurement along $N$-Bell basis.
Since $M_B$, $P^i_{\parallel,a}$ and $P^i_{{\rm anti}-\parallel, a}$
are diagonal with respect to a single basis ($N$-Bell basis), they clearly
commute with each other. Thus, a pre-measurement $M_B$ by Eve along
$N$-Bell basis will in no way change the outcome for
$P^i_{\parallel,a}$ and $P^i_{{\rm anti}-\parallel, a}$.
With any loss of generality, we
can assume that such a pre-measurement is always performed before
the subsequent measurement of
$P^i_{\parallel,a}$ and $P^i_{{\rm anti}-\parallel, a}$.
In other words, we have a classical mixture of $N$-Bell basis
vectors and classical probability theory refering only to the
$N$-Bell basis vectors is, thus, valid.
For this reason, estimation of the number of
singlets as well as confidence levels of such an estimation
can be done by classical statistical theory.~QED.

\section{Our secure QKD scheme}

We remark that the fraction of singlets, $f_s$,
in Proposition~3 has the significance
as being the fraction of uncorrupted qubits in a quantum
communication channel shared between Alice and Bob in the
following situation. Suppose Alice prepares $N$ EPR pairs locally
and, afterwards, sends a member of each pair to Bob via
a noisy quantum channel controlled Eve.
As a result of channel noises and eavesdropping attack,
some of the $N$ EPR pairs may be corrupted.
Proposition~3 gives us a mathematical estimate of the
number of uncorrupted qubits in the actual transmission,
based on the random sampling of a small number of transmitted
signals.

Since quantum error correcting codes (QECCs) exist, it is
tempting to construct a secure QKD scheme by,
first, using the random sampling procedure to estimate the
error rate of the transmission and, second,
using a QECC to correct the appropriate number of
errors. To ensure that the sampling procedure is, indeed,
random, Alice should mix up the test pairs
with the pairs in the actual QECC randomly.

However,
as briefly noted in the Introduction,
the above idea implicitly assumes that the following
conjecture is
true. Let us consider the four error operators
$I$, $\sigma_x$, $\sigma_y$ and $\sigma_z$ for each
quantum signals transmitted. (See appendix~\ref{physics} for notations.)

{\bf Conjecture~4}:
The error rate of a quantum communication channel is
independent of the signals being transmitted.
More precisely, in the current case,
one can safely assign a probability for
each error pattern in analyzing the security issue of
QKD scheme.

While such a conjecture is intuitively
plausible, we are unaware of any rigorous proof for
a general quantum channel. To address this problem, we
prove a related but perhaps weaker result concerning
a teleportation channel. We make use of the
well-known fact that, any quantum signals can always be
transmitted to a quantum communication channel via
teleportation.

\subsection{Teleportation}
\label{ss:tele}

In teleportation \cite{tele}, a quantum signal is transported via
a dual usage of prior ``entanglement'' (i.e. standard EPR pairs
shared between the sender, Alice, and the receiver, Bob) and a classical
communication channel. The quantum signal in Alice's hand
is destroyed by her local measurement, which generates
a classical message.
This message is then transmitted to Bob via a classical
communication
channel.
Depending on the content of this
message, Bob can then re-construct the
destroyed quantum signal by applying one of the unitary transformations
$I$, $\sigma_x$, $\sigma_y$ and $\sigma_z$ to
each of his member of the EPR pairs originally shared with Alice.

Two points are noteworthy. First,
in teleportation the same
prior entanglement is shared by Alice and Bob,
independent of the actual
quantum signal that will subsequently be transported.
Now, since Alice always sends the same standard quantum
signal to Bob during the
prior sharing part of the teleportation
process, the discussion of classical random sampling
theory in subsection \ref{application} can be
applied directly.
Second, the re-construction step in
teleportation, if done with reliable quantum computers,
will not introduce new errors into the quantum system. Indeed,
if Alice and Bob use a {\it noisy} quantum state
shared between them for teleportation,
for each transmitted signal, the three types of
errors $\sigma_x$, $\sigma_y$ and $\sigma_z$
are simply permuted to one another during the re-construction
process. This idea is true even for a quantum
superposition of error patterns and entanglement with
external universe (as specified by the original noisy
quantum state shared between them).

Let us formulate this result mathematically.
Consider the teleportation of a system $\cal S$ consisting of
$N$ qubits from Alice to Bob with the most general mixed
state $\rho_u$.
Without loss of generality, a system decribed by
a mixed state can be equivalently described by a pure state of
a larger system consisting of the original system and an ancilla.
(John Smolin has coined the name ``the Church of the larger Hilbert space''
for this simple but useful observation, which has recently been
extensively used
\cite{sdeutsch,LCbitcom,Mayersbitcom,Lo}.
For instance, the generality of the recent proofs of the
impossibility of bit commitment\cite{LCbitcom,Mayersbitcom}
and one-out-of-two
oblivious transfer\cite{Lo} follows from this idea.)
Applying this idea to our current case, the state of
original system $\cal S$ (plus an ancilla $\cal R$ with which it is
entangled) can be written in the following
form (so-called Schmidt decomposition): 
\begin{equation}
| v \rangle_{\cal RS}=
\sum_m c_m | w_m \rangle_{\cal R} | v_m \rangle_{\cal S},
\label{v}
\end{equation}
where $c_m$ are some complex coefficients, $| w_m \rangle_{\cal R}$ and
$| v_m \rangle_{\cal S}$ are some basis vectors of the two systems $\cal R$
and $\cal S$ respectively. The initial
state $\rho_u$ of the $N$ pairs shared by Alice and Bob can also
be {\it purified} in ``the Church of the larger Hilbert space''
as
\begin{equation}
| u \rangle = \sum_{i_1, i_2 , \cdots, i_N} \sum_j
\alpha_{i_1, i_2 , \cdots, i_N,j} | i_1, i_2 , \cdots, i_N \rangle
\otimes | j \rangle ,
\label{two0}
\end{equation}
where $i_k$ denotes the state of the $k$-th pair and
it runs from $\tilde{0} \tilde{0}$ to $\tilde{1}\tilde{1}$, the
$| j \rangle$'s
form an orthonormal basis
for the environment (or an ancilla
prepared by Eve), and $\alpha_{i_1, i_2 , \cdots, i_N,j}$ are some complex
coefficients.
Each state $| u \rangle$ represents a particular mixed state.
Note that $| u \rangle $ can be re-written as an
entangled sum of a linear superposition of various
error patterns. i.e.,
\begin{equation}
| u \rangle = \sum_{i_1, i_2 , \cdots, i_N} \sum_j
\alpha_{i_1, i_2 , \cdots, i_N,j} (\prod_k \sigma^{(k)}_{i_k}) 
| \Psi^- \rangle^N
\otimes | j \rangle ,
\label{u}
\end{equation}
where $\sigma^{(k)}_{i_k} $ acts on Bob's member of the $k$-th pair as either
$I$, $\sigma_x$, $\sigma_y$ or $\sigma_z$ depending on the
value of $i_k$, and $| \Psi^- \rangle$ denotes an EPR pair.
With such notations, one can prove our main proposition.

{\bf Proposition~5: Invariance of error rate under teleportation.}
In the above notations, suppose the system $\cal S$ (described
by $| v \rangle_{\cal RS}=
\sum_m c_m | w_m \rangle_{\cal R} | v_m \rangle_{\cal S}$ of the combined system
$\cal R$ and $\cal S$ in Eq.~(\ref{v})) is teleported 
using the $N$ pairs shared by Alice and Bob (described by
$| u \rangle $ of the combined system of the $N$ pairs
and Eve's ancilla in Eq.~(\ref{u})). Suppose further that the
classical outcome of Alice's measurements is $\{j_k\}$.
i.e., she
informs Bob to use the operator 
$ \prod_k \sigma^{(k)}_{j_k}$ for the re-contruction process.
Then, Bob's re-constructed state for the
combined system $\cal R$, $\cal S $ and $\cal E$ can be described
by
\begin{equation}
\sum_m c_m | w_m \rangle_{\cal R} \sum_{i_1, i_2 , \cdots, i_N} \sum_j
\alpha_{i_1, i_2 , \cdots, i_N,j}
\left[ \prod_k \left(  \sigma^{(k)}_{j_k}\sigma^{(k)}_{i_k}\sigma^{(k)}_{j_k}
\right)
\right] | v_m \rangle_{\cal S}
\otimes | j \rangle.
\end{equation}

{\it Remark}: The set of complex coefficients 
$c_m  \alpha_{i_1, i_2 , \cdots, i_N,j} $ remain totally
unchanged under teleportation. For each teleportation
outcome labelled by $\{j_k\}$,
the only real change lies in the conjugation action in
the error operator acting on the
subsystem $\cal S$. i.e.,  $\sigma^{(k)}_{i_k} \to 
 \sigma^{(k)}_{j_k}\sigma^{(k)}_{i_k}\sigma^{(k)}_{j_k}$ for
each $k$. (Recall that $\sigma^{(k)}_{j_k}$ is always
its own inverse.) Since under such conjugation
the trivial error operator (i.e., the identity $I$) is invariant
and the three non-trivial error operators $\sigma_x$,
$\sigma_y$ and $\sigma_z$ are permuted to one another,
the error rate of the teleported signal is
exactly the
same as the original $N$ EPR pairs.

{\bf Proof of Proposition~5}: A straightforward exercise in
quantum information theory \cite{tele},
which we will skip here. 

\subsection{Procedure of our secure QKD scheme}

Having established Proposition~5, we now present the
procedure of our secure QKD scheme.

1) Alice prepares $N$ EPR pairs and sends a member of each pair to
Bob through a noisy channel. [In theory, quantum repeaters \cite{Dur} and
two-way schemes for so-called entanglement purification \cite{BB84}
(a generalization of quantum error correcting codes)
could be used in this step. The error rate here can,
therefore, be made to be very small and
the scheme works even for arbitrarily long distances.)

2) Bob publicly announces his receipt of the $N$ quantum signals.

3) Alice {\it randomly} picks $m$ of the $N$ EPR pairs for testing.
She publicly announces her choice to Bob.
For each pair, Alice and Bob randomly pick one of the
three ($x$, $y$, and $z$) axes and perform a measurement
on the two members along it.

4) Alice and Bob publicly announce their measurement outcomes
and use classical sampling theory to estimate
the error rate in the transmission.

{\it Remark}: Proposition~3 allows Alice and Bob to apply classical
sampling theory to the quantum problem at hand to estimate the
error rate of the untested particles. Alice and Bob then
proceed with quantum error correction in the next step.

5) Alice prepares say $R$ EPR pairs and
encodes the $R$ halves of the pairs (i.e., one member from
each pair) by a quantum error correcting code (QECC) into
$N-m$ qubits.

{\it Remark}: The requirement of QECC will be discussed in
subsection~\ref{ss:ftqc}.

6) Alice teleports the $N-m$ qubits to Bob via the remaining $N-m$
pairs that they share.

{\it Remark}: Proposition~5 guarantees the invariance of error
rate under teleportation. So, the estimate done by Alice and Bob in
step 4) remains valid.

7) Alice and Bob perform fault-tolerant quantum computation to
generate a random $R$-bit key by measuring the state of the
$R$ encoded EPR pairs along a prescribed common axis (say the $z$ axis).

\subsection{Fault-tolerant quantum computation}
\label{ss:ftqc}
From Proposition~3 and~5,
it is quite clear that, assuming reliable
local quantum computers, our scheme works perfectly.
However, since local quantum computations may
be imperfect, errors may be generated during the teleportation and
key generation, i.e., steps 6) and 7). One can easily take those local errors into
account by a choice of QECC with generous error-correcting
and fault-tolerant capabilities. The point is that
we have a very specific
and short computation in mind (measurement along $z$ axis only
and no
unitary computation at all). Based on any realistic
error model for quantum computers and concrete choice of
QECC, one can give
a generous upper bound on the number of local errors
due to imperfect quantum computation.
With a fault-tolerant implementation,
the total number of errors in the whole process (transmission,
teleportation and key generation)
can be bounded. Therefore, provided that
our QECC has a sufficiently generous error-correcting
and fault-tolerant capabilities, security is guaranteed.
[To be precise, in step 5), the $R$ EPR pairs
should be prepared fault-tolerantly in an {\it encoded} form
rather than in an unencoded form.]
We remark that, since the required quantum computation here is
much simpler than in \cite{LCqkd}, the present QKD scheme may be
more efficient than the one there.

\section{Concluding Remarks}

In summary, we have presented a simple proof of
the unconditional security of quantum
key distribution, i.e., ultimate security against the most general
eavesdropping attack and the most general
types of noises. Our scheme
allows secure QKD over arbitrarily
long distances, but it requires Alice and Bob to have reliable
quantum computers, which is far beyond current technology.
However, to put things in perspective, all proposed proofs of
security of QKD involve
assumptions (such as
ideal sources) that are beyond current technologies.

Notice that some of the techniques
developed here and in \cite{LCqkd} have widespread applications.
For example, Note 21 of \cite{LCqkd} shows
that teleportation is a powerful technique against the
quantum Trojan Horse attack.
A new application---use random sampling and
random teleportation to prove the feasibility of a general
two-party fault-tolerant quantum computation even in the presence of
eavesdroppers---will be discussed in appendix~\ref{a:two-party}.
In fact, some of the results are applicable even to the case when
Alice and
Bob do not have a quantum computer. A good example
is a quantitative statement on the tradeoff between information
gain and disturbance in BB84 \cite{LCqkd}.

We particularly thank P. W. Shor for inspiring discussions.
Very helpful comments from C. H. Bennett, H. F. Chau, and
John Smolin are also gratefully
acknowledged.

\appendix

\section{Physics Background: Einstein-Podolsky-Rosen pairs}
\label{physics}
The fundamental unit of quantum information is called a quantum
bit or ``qubit''.
Physically, it is often represented by a two-level microscopic system
such as an atom or nuclear spin or a polarized photon.
Mathematically, a pure quantum state of a qubit
simply given by a unit vector in a two-dimensional Hilbert space
${\cal H}^2$:
Let us consider any basis $| 0\rangle $ and $| 1 \rangle$.
A single qubit in a pure state can be in any superposition of
the two basis vectors, i.e., $ a | 0 \rangle + b | 1 \rangle$
where $a$ and $ b$ are complex coefficients with the normalization
$| a|^2 + | b|^2 = 1$.
A pair of qubits is described by a unit vector in
the tensor product space ${\cal H}^4 = {\cal H}^2 \times {\cal H}^2$.
with the basis states $| 00 \rangle$,
$| 01 \rangle$, $| 10 \rangle$ and $| 11 \rangle$.
Consider the state $| \Psi^-\rangle
= \sqrt{ 1/2} (| 01 \rangle - | 10 \rangle)$.
The important point to note is
that it is impossible to re-write $| \Psi^-\rangle$ into the form
of a direct product
$|u\rangle \otimes | v\rangle$.
The state $| \Psi^-\rangle$ is called {\it entangled} because it is impossible to
assign a definite state to the individual subsystems.
$| \Psi^-\rangle$ is called an Einstein-Podolsky-Rosen (EPR)
pair.

It is common to write $ a | 0 \rangle + b | 1 \rangle$ also as
a column vector $\left( \begin{array}{c}
         a \\
         b
       \end{array} \right)$.
The non-trivial error operators (or Pauli matrices) are defined as $\sigma_x = 
\left( \begin{array}{cc}
         0 & 1 \\
         1 & 0 
       \end{array} \right)$,
$\sigma_y = 
\left( \begin{array}{cc}
         0 & -i \\
         i & 0 
       \end{array} \right)$,
and
$\sigma_z = 
\left( \begin{array}{cc}
         1 & 0 \\
         0 & -1 
       \end{array} \right)$.

\section{Bell basis}
\label{a:Bell}
The basis vectors of the
Bell basis are $\Psi^{\pm} $ and $\Phi^{\pm}$, where
\begin{equation}
\Psi^{\pm}= { 1 \over \sqrt{2}}
( | \uparrow \downarrow \rangle \pm | \downarrow \uparrow \rangle)
\end{equation}
and
\begin{equation}
\Phi^{\pm}= { 1 \over \sqrt{2}}
( | \uparrow \uparrow \rangle \pm | \downarrow \downarrow \rangle).
\end{equation}
With the convention in
Ref. \cite{BDSW},
Bell basis vectors are represented by two classical bits:
\begin{eqnarray}
\Phi^+ & =& \tilde{0}\tilde{0}, \nonumber \\
\Psi^+ & =& \tilde{0}\tilde{1}, \nonumber \\
\Phi^- & =& \tilde{1}\tilde{0}, \nonumber \\
\Psi^- & =& \tilde{1}\tilde{1}.
\label{bellstate}
\end{eqnarray}
\noindent
Since Bell basis vectors are highly entangled, one should not
think of them as direct product states.

\section{Two-party fault-tolerant quantum computation in the
presence of an eavesdropper}
\label{a:two-party}
Here we show that
random sampling and
random teleportation can be used to prove the feasibility of a general
two-party fault-tolerant quantum computation even in the presence of
eavesdroppers. This may look hard because the usual
requirements of fault-tolerant
quantum computation demand that the errors
of different signals are independent and that
the error rate for each error to happen is
smaller than some threshold value. In contrast, an eavesdropper
can introduce collective noises into the system.

{\bf Proposition~6}:
In the large $N$ limit, the procedure in Proposition~3 can be used to
establish that, with a very high
confidence level, the error rates of
the transmitted signals are well below the threshold value
required for a general
fault-tolerant quantum computation
and that the error rates for
different signals are essentially {\it independent}.

{\bf Proof}:
Suppose $N$ quantum signals are teleported via $N$ EPR pairs
such that each signal is teleported by a {\it random}
pair (without replacement, of course)
chosen by Alice and Bob.
By Propositions~3 and~5,we can apply classical
sampling theory to our current quantum problem.
Now, since the signals are {\it randomly} sampled,
in the large $N$ limit of classical sampling theory, they
have {\it identical} and {\it independent} error probabilities.
Therefore, by random sampling and random teleportation,
Alice and Bob can establish confidence levels for
the smallness and independence of the error rates
of different signals, thus allowing
subsequent fault-tolerant quantum computations.

{\it Remark}: The fact that our claim is valid is not that surprising.
In classical computation, it is natural and often
implicit to assume that reliable
two-party classical computation (such as authentication)
can be performed with imperfect
computing components and noisy classical communication
channels controlled by an eavesdropper.
It is only natural that the same assumption can be made for
two-party quantum computation.

\end{document}